%% file: main_text.tex
\documentclass[%
 reprint,
superscriptaddress,
 amsmath,amssymb,
 aps,
 prl,
floatfix,
]{revtex4-2}

\usepackage{url}
\usepackage{hyperref}
\usepackage{longtable}
\usepackage{bbm}
\usepackage{times}
\usepackage{nicefrac}
\usepackage{amsfonts}
\usepackage[ansinew]{inputenc}
\usepackage{epsfig}
\usepackage{graphicx}
\usepackage{color}
\usepackage{bm}
\usepackage{multirow}
\usepackage{mathrsfs}
\usepackage{revsymb}
\usepackage{amssymb}
\usepackage{amsmath}
\usepackage{mathtools}
\usepackage{slashed}
\usepackage{dcolumn}
\usepackage{textcomp}
\usepackage{upgreek}

\usepackage{lineno,hyperref}
\modulolinenumbers[5]

\usepackage{amsmath}
\usepackage{xspace}

\usepackage{graphicx}
\usepackage{dcolumn}
\usepackage{bm}

\usepackage{subfigure}% bold math

\newcommand{\pentatrap}{\textsc{Pentatrap}\xspace}

\newcommand{\shiptrap}{\textsc{Shiptrap}\xspace}

\newcommand{\orcid}[1]{\href{https://orcid.org/#1}{\includegraphics[scale=0.06]{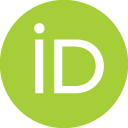}}}

\begin{document}

\title{Direct \textit{Q}-Value Determination of the $\beta^-$~Decay of $^{187}$Re}

\author{P. Filianin\orcid{0000-0002-7797-9873} }
\email{filianin@mpi-hd.mpg.de}
\affiliation{Max-Planck-Institut f{\"u}r Kernphysik, 69117 Heidelberg, Germany}

\author{C.~Lyu\orcid{0000-0001-9444-6385}}
\author{M.~Door\orcid{0000-0003-2455-3198}}
\author{K.~Blaum\orcid{0000-0003-4468-9316}}
\affiliation{Max-Planck-Institut f{\"u}r Kernphysik, 69117 Heidelberg, Germany}

\author{W.J.~Huang\orcid{0000-0002-5553-3942}}
\affiliation{Advanced Energy Science and Technology Guangdong Laboratory, Huizhou 516003, China}

\author{M.~Haverkort\orcid{0000-0002-7216-3146}}
\affiliation{Institute for Theoretical Physics, Heidelberg University, 69120 Heidelberg, Germany}

\author{P.~Indelicato\orcid{0000-0003-4668-8958}}
\affiliation{The University of Sorbonne, 75006 Paris, France}

\author{C.H.~Keitel\orcid{0000-0002-1984-1470}}
\author{K.~Kromer\orcid{0000-0002-6057-356X}}
\author{D.~Lange\orcid{0000-0002-4559-6739}}
\affiliation{Max-Planck-Institut f{\"u}r Kernphysik, 69117 Heidelberg, Germany}

\author{Y.N.~Novikov}
\affiliation{Department of Physics, St Petersburg State University, St Petersburg 198504, Russia}
\affiliation{NRC ``Kurchatov Institute'' - Petersburg Nuclear Physics Institute, 188300 Gatchina, Russia}

\author{A.~Rischka\orcid{0000-0003-2523-5933}}
\affiliation{ARC Centre for Engineered Quantum Systems, School of Physics, The University of Sydney, NSW 2006, Australia}

\author{R.X.~Sch\"ussler\orcid{0000-0003-3423-0982}}
\author{Ch.~Schweiger\orcid{0000-0002-7039-1989}}
\author{S.~Sturm\orcid{0000-0002-0309-5311}}
\affiliation{Max-Planck-Institut f{\"u}r Kernphysik, 69117 Heidelberg, Germany}

\author{S.~Ulmer\orcid{0000-0002-4185-4147}}
\affiliation{Ulmer Fundamental Symmetries Laboratory, RIKEN, Wako, Saitama 351-0198, Japan}

\author{Z.~Harman}
\author{S.~Eliseev\orcid{0000-0003-0210-7082}}
\affiliation{Max-Planck-Institut f{\"u}r Kernphysik, 69117 Heidelberg, Germany}

\date{\today}

\begin{abstract}
The cyclotron frequency ratio of $^{187}\mathrm{Os}^{29+}$ to $^{187}\mathrm{Re}^{29+}$ ions was measured with the Penning-trap mass spectrometer \pentatrap. The achieved result of $R=1.000\:000\:013\:882(5)$ is to date the most precise such measurement performed on ions. Furthermore, the total binding-energy difference of the 29 missing electrons in Re and Os was calculated by relativistic multiconfiguration methods, yielding the value of $\Delta E = 53.5(10)$\,eV. Finally, using the achieved results, the mass difference between neutral $^{187}$Re and $^{187}$Os, i.e., the $Q$~value of the $\beta^-$~decay of $^{187}$Re, is determined to be 2470.9(13)\,eV.
\end{abstract}

\maketitle

At 2.5 keV, the $\beta^-$~decay of $^{187}$Re has the smallest decay energy ($Q$ value) among beta transitions between nuclear ground states. By investigating its $\beta$~spectrum, multiple aspects of fundamental physics can be studied. Among them are the determination of the neutrino rest mass \cite{Nucciotti2, Arnaboldi-2003, formaggio2021direct}, the search for hypothetical sterile neutrinos \cite{Shrock, Adhikari-2017}, weak interaction symmetry tests \cite{Glick-Magid}, and a test of low-energy electron-electron interactions \cite{Nucciotti2, Gatti-1999}. 

The $Q$ value is the key parameter for the description of the spectrum. It can be deduced from the $\beta^-$ spectrum itself via the Kurie plot \cite{Kurie-plot}. However, if the experimentally measured spectrum is affected by systematic effects, and/or if the shape of the spectrum is not accurately known, the extracted $Q$ value  may significantly deviate from the ``true'' one. High-precision Penning-trap mass spectrometry (PTMS) provides a direct and independent way for the determination of the $Q$ value by measuring the mass difference between the parent and daughter nuclides. In order to examine possible effects, which can affect the shape of the spectrum, the directly measured $Q$~value should be known with an uncertainty similar to the energy resolution of the detector, which for modern cryogenic microcalorimeters (CM) is on the order of a few eV in the energy range of a few keV.

Because of a long half-life of 41.2\,Gy, the $\beta^-$ decay of neutral $^{187}$Re is usually recorded in a solid state where a large concentration of $^{187}$Re atoms can be found. The electron emitted from such a decay process is most likely to be trapped in the corresponding material, and the singly charged $^{187}\mathrm{Os}^{+}$ ion will be eventually neutralized. As a result, the kinetic energy (except that from the antineutrino) released from the $\beta^-$ decay can be transformed into heat that can be measured by CM \cite{Alessandrello-1999, Galeazzi-2000, Arnaboldi-2003}. Thus, the mass difference $\Delta M$ of neutral $^{187}$Re and $^{187}$Os atoms can be considered approximately equal to the $Q$ value of the $\beta^-$~decay of $^{187}$Re in solid state, which can be inferred from  the endpoint energy $E_0$ recorded by the CM. The difference of up to a few eV between the $Q$ value and $\Delta M$ can arise from solid state effects that may shift binding energies for $^{187}$Re and $^{187}$Os atoms in a rhenium crystal. Thus, a direct determination of the $\Delta M$ by PTMS will probe the validity of the theoretical model employed to describe the  $\beta^-$-decay spectrum, and might point at possible systematic effects inherent in the CM technique.\\

The endpoint energy $E_0$  recommended  by the atomic mass evaluation (AME) \cite{AME_16a} is 2466.7(16)~eV. The only direct measurement of the $\Delta M$  was carried out with \shiptrap, yielding $\Delta M = 2492(33)$\,eV \cite{Nesterenko}. A necessary improvement toward an eV level in the precision of $\Delta M$ determination is required. It has become possible with the development of the experiment \pentatrap \cite{repp2012pentatrap,roux2012trap,Xenon, Re_Nature} and elaborate calculations of the electron binding energies  in Re and Os atoms.\\

\begin{figure}
\includegraphics[width=1\linewidth]{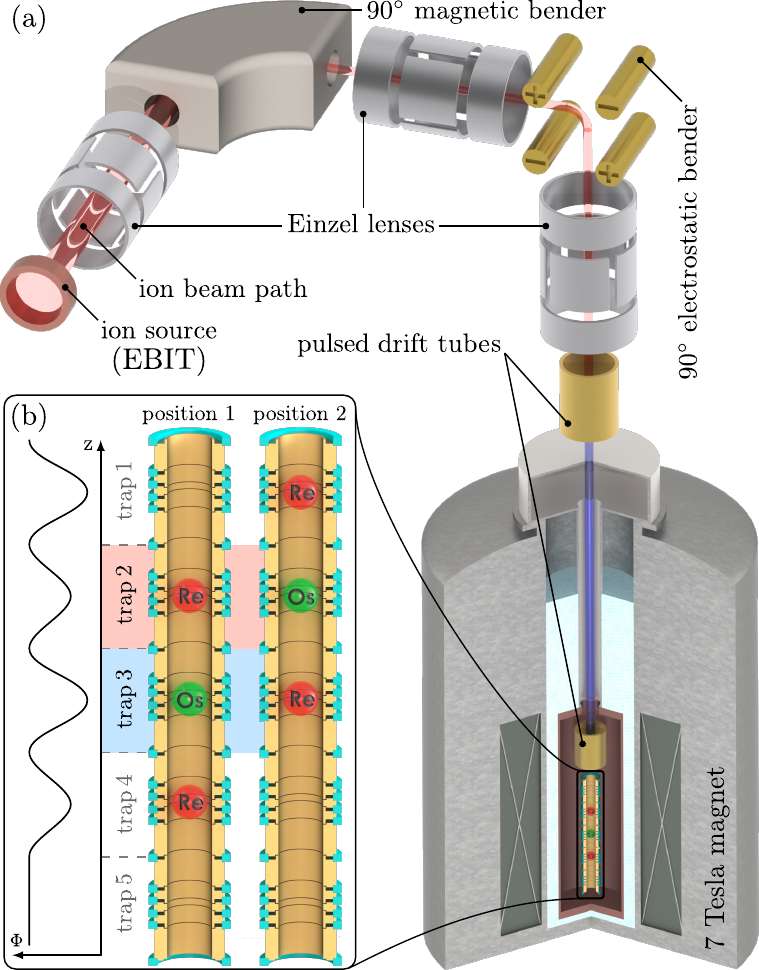}
\caption{(a) Schematic of the \pentatrap setup (not to scale). The ions are produced in an external ion source (electron beam ion trap, or EBIT) with a kinetic energy of 6.5 keV/$q$, get $m/q$ separated with a magnetic bender, slowed down with two pulsed drift tubes to a few eV/$q$ energy and finally  captured in the Penning traps hosted in a superconducting magnet. (b) Ion positions inside the traps which are sequentially repeated during the measurement. The electric potential $\Phi$ along the trap axis is depicted on the left side. For details see text.}
\label{fig:beamline}
\end{figure}

In this Letter, we report on a mass-difference $\Delta M$ measurement of $^{187}$Re and $^{187}$Os with an uncertainty of 1.3~eV. For this, the measurement of the cyclotron-frequency ratio $R$ of $^{187}\mathrm{Os}^{29+}$ to $^{187}\mathrm{Re}^{29+}$ ions and the theoretical calculation of the binding energy difference $\Delta E=E^{29+}_{\mathrm{Re}}-E^{29+}_{\mathrm{Os}}$ are combined. Here, $E^{29+}_{\text{Re}}$ and $E^{29+}_{\mathrm{Os}}$ are the binding energy of the 29 missing electrons in Re and Os, respectively. In PTMS, the cyclotron frequencies $\nu_c = \frac{1}{2\pi}\frac{q}{m} B$ of ions with charge $q$ and mass $m$ in a uniform magnetic field $B$ are compared. The ion's cyclotron frequency is determined by measuring the eigenfrequencies $\nu_+$, $\nu_-$ and $\nu_z$ of the ion in a Penning trap and by applying  the invariance theorem  $\nu_c^2 = \nu_-^2 + \nu_z^2 + \nu_+^2$ \cite{Brown}. The $\Delta M$ is determined as 
\begin{equation}
\label{eq:Q-value}
\Delta M = \left[ m(^{187}\mathrm{Os}) - 29m_e - E^{29+}_\mathrm{Os} \right] \left[ R-1 \right] + \Delta E,
\end{equation}
where $m_e$ is the electron mass \cite{SSturm} and $m(^{187}\mathrm{Os})$ is the mass of the neutral $^{187}\mathrm{Os}$ atom. The masses are given in energy units. Since $(R-1) \sim 10^{-8}$, the contributions of the $E^{29+}_{\text{Os}}$ term and of the uncertainties of $m({}^{187}\text{Os})$ and $m_e$ are much smaller than that from $R$ and $\Delta E$ terms and, thus, can be neglected. The value of $m(^{187}\mathrm{Os})$ is taken from \cite{AME_16a}. 

Here just an overview of the measurement procedure and a brief analysis of the obtained data are presented. A comprehensive description of the data analysis will be given in  \cite{Re-analysis}. 

Fig.~\ref{fig:beamline} is a schematic of the \pentatrap setup. For the production of highly charged rhenium and osmium ions, their volatile organic compounds are alternately injected into a room-temperature commercial electron beam ion trap (EBIT) \cite{zschornack2010status}. After 200\,ms of charge-breeding time in the EBIT, a few $\upmu$s-long bunches of ions are ejected with 6.5~keV/$q$ kinetic energy, charge-to-mass ratio separated in a $90^\circ$ magnetic bender, slowed down with two pulsed drift tubes to a few eV energy, and finally captured in the Penning traps situated inside a 7\,Tesla superconducting magnet. \pentatrap has five identical cylindrical Penning traps that, with the associated detection electronics, are cooled to liquid helium temperature. An ultrastable voltage source is used to create the traps' potential \cite{Starep}. Various measures are undertaken to stabilize the magnetic field of the magnet. Presently,  trap\,2 and trap\,3 are used to measure the ions frequencies, whereas the remaining three traps serve for ion storage. Trap\,1 and the above mounted pulsed drift tube are used to load ions. In order to load just a single ion into trap\,1, the intensity of the ion beam ejected from the EBIT is reduced such that  approximately every tenth loading attempt an ion reaches trap\,1. After every loading attempt the content of trap\,1 is transported to trap\,2 for its identification and preparation. The preparation consists of the reduction of the ion's motional amplitudes with the resistive cooling technique \cite{Brown}. Great care is taken to make sure that there are no other ions in the trap except for a single ion of interest. Following this preparation procedure, the ion can be transported to any of  traps 3, 4 or 5  with a subsequent loading of the next ion.

\begin{figure*}
\includegraphics[width=1\textwidth]{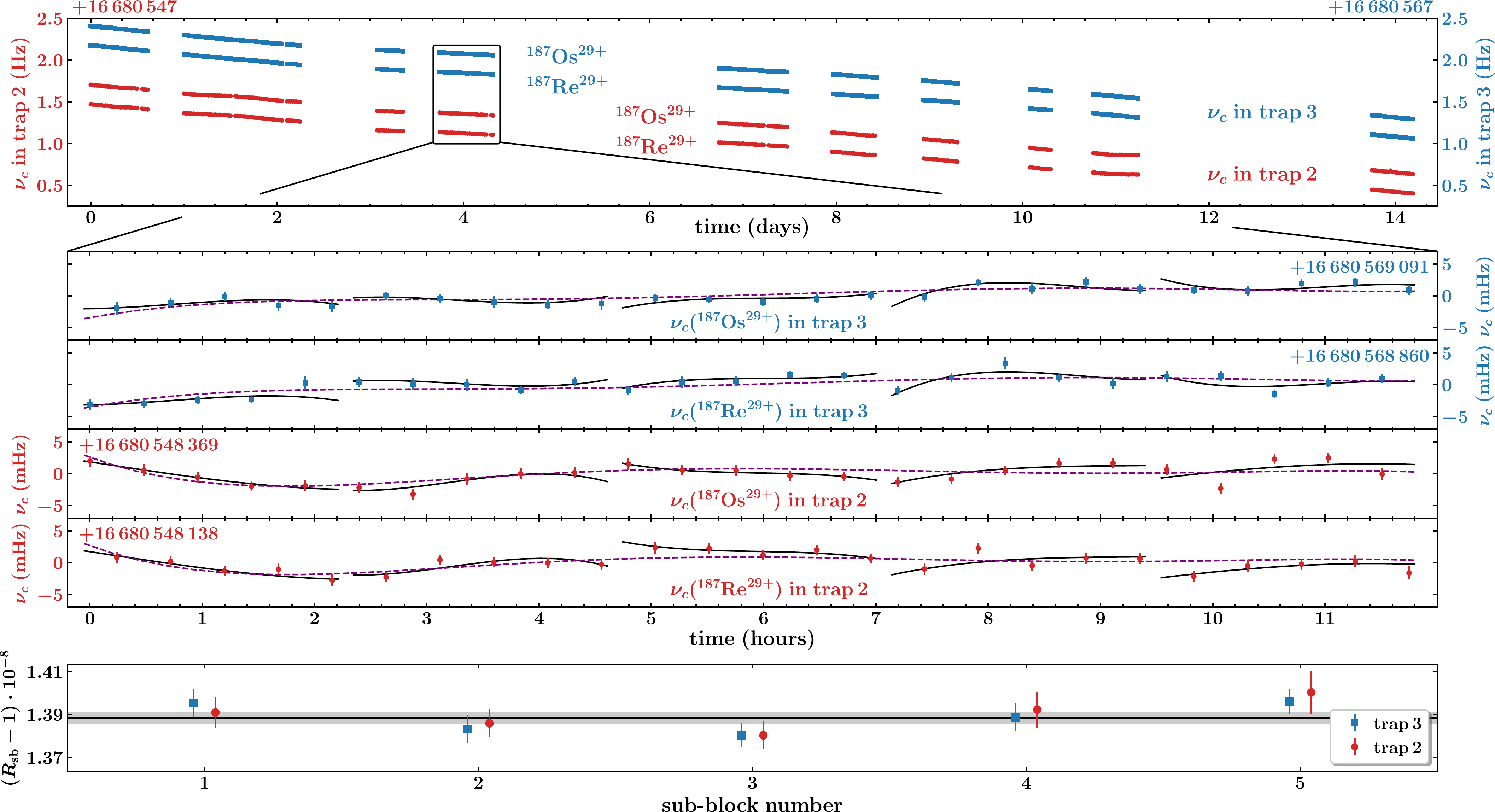}
\caption{Top: the cyclotron frequencies of $^{187}\mathrm{Os}^{29+}$ and $^{187}\mathrm{Re}^{29+}$ ions in trap\,2 (in red, left ordinate) and trap\,3 (in blue, right ordinate) acquired over the entire measurement campaign, giving, in total, about 1100 $\nu_c$ data points. Middle four plots: enlargement of data block 4 plotted at the top. The interval is divided into five sub-blocks and the simultaneous fit of the third-order polynomials is applied to each sub-block for both traps (solid curves). Alternatively, a fifth-order polynomial was found to be the most reasonable model for the entire data block using the corrected Akaike information criterion as a model quality estimator (dashed purple curves). A common drift of $-2.4$ mHz/h is subtracted for the sake of clarity of the scattered data points. Bottom: the corresponding cyclotron frequency ratios $R_\mathrm{sb}$ determined for each sub-block for both traps. The black line and the gray shaded band are the weighted mean of the $R_\mathrm{sb}$ values for both traps and its statistical uncertainty, respectively.}
\label{fig:freqs}
\end{figure*}

The ions' eigenfrequencies are measured as follows. Since the magnetron frequency $\nu_-$ is almost exactly the same for ions of both nuclides, it is  measured  once with a moderate uncertainty. Thus, the  procedure reduces to a simultaneous measurement of  the reduced cyclotron frequency $\nu_+$  with the phase-sensitive pulse-and-phase technique \cite{PnP2, PhysRevA.41.312} and the axial frequency $\nu_z$ with the ``single-dip'' method \cite{feng1996tank}. The measurement is performed synchronously on two ions situated in two traps, giving an additional flexibility in the analysis of the acquired data. A set of 3 ions is loaded into the starting position 1 [Fig.~\ref{fig:beamline}(b)]. The first and the last ion is $^{187}\mathrm{Re}^{29+}$; the middle one is $^{187}\mathrm{Os}^{29+}$. The $\nu_+$ and $\nu_z$ frequencies of the ions in trap\,2 and trap\,3 are then  measured for approximately 15\,min. Afterwards, the ions are moved by consecutive adiabatic transport into the neighboring traps [position 2 in Fig.~\ref{fig:beamline}(b)] with subsequent measurements of their eigenfrequencies for again about 15 min. This procedure is repeated until the measurement is stopped. The measurements take place at nights and on weekends when the stray magnetic field and the temperature fluctuations in the laboratory are minimal. Typically, between the night measurements the traps are emptied and a new set of three ions is loaded. In this way, the systematic frequency shifts due to a potential presence of unwanted additional ions circling on large orbits in the traps are avoided. A measurement with the same set of ions is called a data block. The total measurement campaign consists of 10 such data blocks (see Fig.~\ref{fig:freqs}, top).

\begin{figure*}
\includegraphics[width=1\linewidth]{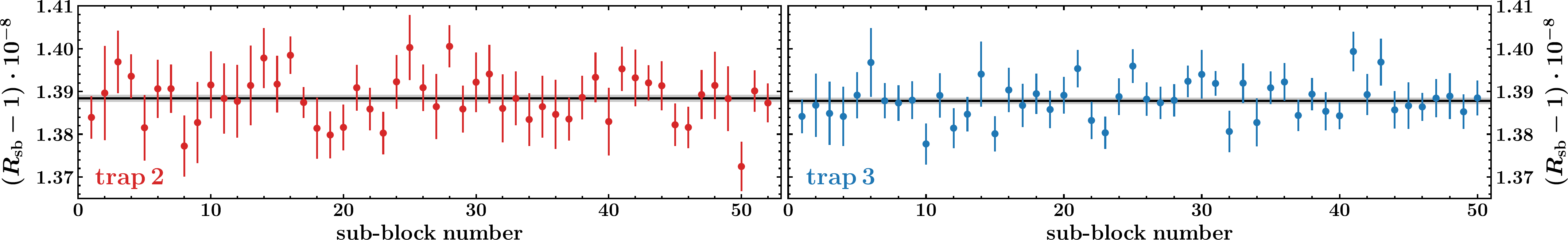}
\caption{The cyclotron frequency ratios $R_\mathrm{sb}$ determined over the entire measurement campaign for trap\,2 (left) and trap\,3 (right). The ratios are calculated by fitting polynomials of the third order to the sub-blocks of data, as shown in Fig.~\ref{fig:freqs}, middle. The weighted mean value and its statistical uncertainty for each trap are represented by the black line and the gray shaded band, respectively.}
\label{fig:ratios}
\end{figure*}

Several methods are employed for the $\nu_c$ data analysis to provide a reliable assessment of the statistical and systematic uncertainties. Although based on method-specific assumptions and approximations, all methods yield statistically equal frequency ratios with similar uncertainties. The mean between the values obtained using the applied analysis methods $\overline{R}_\mathrm{final}$ is considered the final cyclotron frequency ratio to be used for the determination of the $\Delta M$. Since the complete detailed data analysis will be published soon in a dedicated paper \cite{Re-analysis}, only the so-called ``polynomial'' method, which has already become conventional in the PTMS community \cite{FSU_POLY, Xenon, Re_Nature}, is presented here. The method is based on the reasonable assumption that the time dependence of the magnetic field drift and, hence, of the $\nu_c$ drift can be approximated by a polynomial. Thus, the $\nu_c$ frequency drift of Os and Re ions is approximated by two polynomials that differ only by a coefficient of proportionality $R$. The frequency ratios $R$ are then obtained from the simultaneous fit of two polynomials to the chosen $\nu_c$ dataset. 

Two approaches have been used within the polynomial method. The first approach is based on the division of each data block into two-hour long sub-blocks. The polynomials of third order are then fitted to the $\nu_c$ dataset of each sub-block (Fig.~\ref{fig:freqs} middle), thus yielding a sub-block frequency ratio $R_\mathrm{sb}$ (Fig.~\ref{fig:freqs} bottom). The final frequency ratio and its internal and external uncertainties \cite{Birge} are determined by a weighted average of all $R_\mathrm{sb}$ values from trap\,2 and trap\,3 (Fig.~\ref{fig:ratios}). The final uncertainty is the larger of the internal and external uncertainties. This approach has a certain degree of freedom in the choice of the polynomial order and  the size of the sub-block. This is accounted for by a variation of the sub-block size and the polynomial order with a subsequent adjustment of the uncertainty of the final frequency ratio. For the second approach, the entire data block is subject to the polynomial fit. The polynomial order is chosen on the basis of the corrected Akaike information criterion \cite{Hurvich}.

Ultimately, the final cyclotron frequency ratio is $\overline{R}_\mathrm{final} = 1.000\,000\,013\,882(5)(1)$, where the values in parentheses are the statistical and systematic uncertainties, respectively. Hereinafter, the digits in parentheses are the one-standard-deviation uncertainty in the last digits of the given value. The systematic uncertainty is dominated by the uncertainty in the determination of the ion cyclotron frequency due to a nonlinear deviation of the pulse-and-phase readout phase from the real phase of the ion cyclotron motion. Since the fractional mass difference between the $^{187}\mathrm{Re}^{29+}$ and $^{187}\mathrm{Os}^{29+}$ ions is only $10^{-8}$, all other contributions to the systematic uncertainty in the determination of the final ratio due to, for example, image charge shift, relativistic shift, the anharmonicity of the trap potential, and the inhomogeneity of the magnetic field are well below $10^{-12}$ and hence can be neglected \cite{Re-analysis}.

%%%%%%%%%%%%%%%%%%%%%%%%%%%% Theory part %%%%%%%%%%%%%%%%%%%%%%%%%%%%%%%

In order to determine the $\Delta M$ from $\overline{R}_\mathrm{final}$, the total binding energy of the 29 outermost electrons $E^{29+}_\mathrm{Re}$ in the Re atom and $E^{29+}_\mathrm{Os}$ in the Os atom, and their difference $\Delta E$, is subject to theoretical calculations. For this, two fully relativistic multiconfiguration methods are employed. The first one is the multiconfiguration Dirac-Hartree-Fock method (MCDHF) \cite{Grant1970, Desclaux1971} and its combination with the Brillouin-Wigner many-body perturbation theory~\cite{Kotochigova2007, GRASP2018}. While the ground state of the $\mathrm{Re}^{29+}$ ion is a simple configuration [Kr]$4d^{10}$~${}^1S_0$, the neutral Re atom is in the [Xe]$4f^{14}5d^{5}6s^{2}$~${}^6 S_{5/2}$ electronic state. The Os ion and atom have an additional electron compared to their Re counterparts, thus their ground states are the [Kr]$4d^{10}4f$~${}^2F_{5/2}$ and [Xe]$4f^{14}5d^{6}6s^{2}$~${}^5 D_{4}$ configurations, respectively. Within the MCDHF scheme, the many-electron atomic state function is given as a linear combination of configuration state functions (CSFs) with a common angular momentum, magnetic and parity quantum numbers. The CSFs are constructed as $jj$-coupled Slater determinants of one-electron orbitals. Using the GRASP2018 code package~\cite{GRASP2018}, we systematically expand the active space of virtual orbitals used in the generation of CSFs by electron exchanges to monitor the convergence of the binding energy differences, and to assess their calculational uncertainties. The set, consisting of 4 to 5 million CSFs, is generated with single and double electron exchanges from the $3p$--$6s$ states up $9h$, with the virtual orbitals optimized layer by layer. By doing so, we obtain $E^{29+}_\mathrm{Re} = 10894.5(259)$~eV and $E^{29+}_\mathrm{Os} = 10947.9(246)$~eV, with the electron correlation effects contributing tens of eVs.  Nevertheless, correlation terms and errors largely cancel in the binding energy difference $\Delta E = E^{29+}_\mathrm{Re} - E^{29+}_\mathrm{Os}$ due to the similarities of the two atoms and ions. Thus, the MCDHF method yields a rather accurate value of $\Delta E_\mathrm{MCDHF} = 53.4(10)$~eV.

A check with an independent MCDHF code \cite{indelicato2005}, using a more limited set of configurations but with all orbitals fully relaxed and with all-order Breit interaction in the correlation evaluation (see, e.g., \cite{santos2006} and references therein) gives a similar value $\Delta E_\mathrm{BMCDHF} = 53.9(15)$\,eV. It is worth noticing that this result is strongly dominated by the Coulomb interactions.

In another set of calculations performed with the {\textsc{Quanty}\xspace} package \cite{Haverkort2016}, the Hilbert space is spanned by multi-Slater-determinant states constructed from relativistic Kohn-Sham single-electron orbitals, obtained from the density functional code FPLO \cite{Koepernik1999}, using the local spin density functional \cite{Perdew1992}. On this basis, the Dirac-Coulomb-Breit Hamiltonian is solved. In order to make the calculations tractable,  the number of Kohn-Sham orbitals is restricted to a finite set. Within the resulting large but finite Hilbert space,  the determinant states are selected by including only  determinants with a significant contribution to the state of interest \cite{Haverkort2014}.  The basis set is  converged such that the variance of the energy reaches the required accuracy, which was set to $10^{-6}$. To benefit from error cancellation, the difference in binding energies is calculated as $\Delta E = \Delta E^0 - \Delta E^{29+}$, where $\Delta E^0$ and $\Delta E^{29+}$ are the energy differences between Os and Re in their neutral and 29-fold ionized ground states, respectively. Orbitals up to $n_\mathrm{max} = 7$ are included. To extrapolate to $n_\mathrm{max} = \infty$,  results for calculations with $n \le 5$, $n \le 6$, and $n \le 7$ are compared. By successively increasing the number of configurations, the binding energy difference and error estimate are extrapolated to $\Delta E_\mathrm{Quanty} = 53.4(13)$~eV.

The final value $\Delta E_{\mathrm{final}}=53.5(10)$~eV is obtained from the averaging  the theoretical results $\Delta E_\mathrm{MCDHF}$, $\Delta E_\mathrm{BMCDHF}$ and $\Delta E_\mathrm{Quanty}$, weighted with their inverse uncertainties. Since the employed theoretical methods are not fully independent, the 1.0-eV lowest uncertainty among the three theoretical uncertainties is used for $\Delta E_{\text{final}}$. 

%%%%%%%%%%%%%%%%%%%%%%%%%%%%%%%%%%%%%%%%%%%%%%%%%%%%%%%%%%%%%%%%%%%%%%%%%%%%%%%%%%%%

\begin{figure}
\includegraphics[width=1\linewidth]{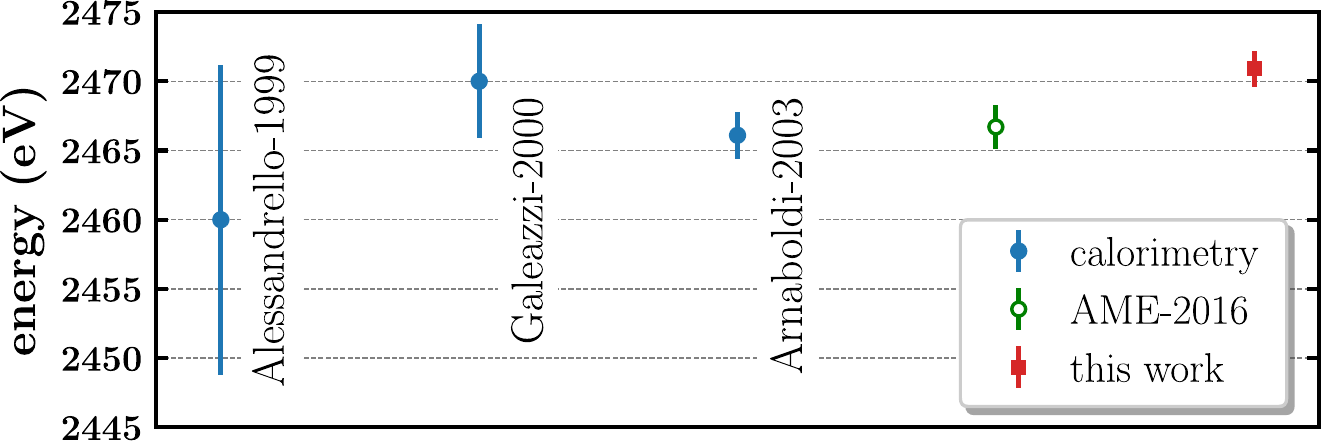}
\caption{The blue filled circles are endpoint energy values $E_0$ obtained from the analysis of the $\beta^-$~decay of $^{187}$Re  in Alessandrello \cite{Alessandrello-1999}, Galeazzi \cite{Galeazzi-2000}, Arnaboldi \cite{Arnaboldi-2003}. The green empty circle is the evaluated $Q$ value from AME \cite{AME_16a}. The red filled square is the $\Delta M$ (in energy units) determined in this work.} 
\label{fig:Q-values}
\end{figure}

\begin{table}[t]
    \caption{\label{tab:results} The results achieved in this work. The digits in parentheses are the one-standard-deviation uncertainty in the last digits of the given value. For details see text.}
    \begin{ruledtabular}
    \begin{tabular}{lr}
        
        $\overline{R}_\mathrm{final}$  & $1.000\:000\:013\:882(5)(1)$   \\[0.1cm]
        $m(^{187}\mathrm{Os}^{29+}) - m(^{187}\mathrm{Re}^{29+})$ & $2\:417.4(9)$~eV \\[0.1cm]
        $\Delta E_\mathrm{MCDHF}$ & 53.4(10)~eV \\[0.1cm]
        $\Delta E_\mathrm{BMCDHF}$ & 53.9(15)~eV \\[0.1cm]
        $\Delta E_\mathrm{Quanty}$ & 53.4(13)~eV \\[0.1cm]
        \hline \\[-0.2cm]
        $\Delta E_\mathrm{final}$ & 53.5(10)~eV \\[0.1cm]
        $\Delta M$ & $2\:470.9(13)$~eV \\[0.1cm]
        
    \end{tabular}
    \end{ruledtabular}
\end{table}

Finally, with the experimentally measured cyclotron frequency ratio $\overline{R}_\mathrm{final}$ and the theoretically calculated electron binding energy differences $\Delta E_\mathrm{final}$, the mass difference between neutral $^{187}$Re and $^{187}$Os is determined to be 2470.9(13)~eV (see Table \ref{tab:results}), which is 25 times more accurate than the previous \shiptrap result. Fig.~\ref{fig:Q-values} shows the comparison between the three most precise endpoint energy values $E_0$ obtained  by different groups using cryogenic microcalorimetry, the evaluated $Q$~value from AME, and the $\Delta M$ obtained in the present work. Within two combined standard deviations, the value obtained in the present work agrees with the value of 2466.7(16)\,eV derived by AME \cite{AME_16a}.  The small difference between the values might be caused by solid state effects in rhenium crystal and thus spurs the need for a more precise $Q$~value or $\Delta M$ determination by both techniques. Although further high-statistical acquisition of the $^{187}$Re $\beta^-$ spectrum  faces certain difficulties, mostly related to demanding technical requirements to the cryogenic microcalorimetric technique, the obtained result may stimulate the resumption of the activities on the accurate $^{187}$Re decay spectra acquisition for the benefit of multiple researches in fundamental physics. \\

This work is part of and funded by the Max Planck Society, the Deutsche Forschungsgemeinschaft (DFG, German Research Foundation) - Project-ID 273811115 - SFB 1225, and by the DFG Research UNIT FOR 2202. P.I. acknowledges partial support from NIST. Laboratoire Kastler Brossel (LKB) is ``Unit\'e Mixte de Recherche de Sorbonne Universit\'e, de ENS-PSL Research University, du Coll\`ege de France et du CNRS n$^\circ$ 8552''. Y.N. thanks Russian Minobrnauki for support within the project n$^\circ$ 10 of Russian-German collaboration. P.I., Y.N. and K.B. are members of the Allianz Program of the Helmholtz Association, contract n$^\circ$ EMMI HA-216 ``Extremes of Density and Temperature: Cosmic Matter in the Laboratory''. This project has received funding from the European Research Council (ERC) under the European Union's Horizon 2020 research and innovation program under grant agreement No. 832848 - FunI. Furthermore, we acknowledge funding and support by  the International Max Planck Research School for Precision Tests of Fundamental Symmetries (IMPRS-PTFS) and by  the Max Planck PTB RIKEN Center for Time, Constants and Fundamental Symmetries.

%\printbibliography
\input{main_text.bbl}

\end{document}

%% file: main_text.bbl
%apsrev4-2.bst 2019-01-14 (MD) hand-edited version of apsrev4-1.bst
%Control: key (0)
%Control: author (8) initials jnrlst
%Control: editor formatted (1) identically to author
%Control: production of article title (0) allowed
%Control: page (0) single
%Control: year (1) truncated
%Control: production of eprint (0) enabled
%